# A novel approach towards selective bulk synthesis of delaminated graphenes in an electric arc

Soumen Karmakar[1], Naveen V Kulkarni[1], Ashok B Nawale[1], Niranjan P Lalla[2], Ratikant Mishra[3], V G Sathe[2], S V Bhoraskar[1, 5] and A K Das[4]

[1]Department of Physics, University of Pune, Pune 411007, India

[2]UGC-DAE Consortium for Scientific Research, University Campus, Khandwa Road, Indore 452017, India

[3]Chemistry Division, Bhabha Atomic Research Centre, Trombay, Mumbai 400085, India

[4]Laser and Plasma Technology Division, Bhabha Atomic Research Centre, Trombay, Mumbai 400085, India

[5]Author to whom any correspondence should be addressed

E-mail: svb@physics.unipune.ernet.in

**Abstract**

Here we demonstrate the selective bulk scale synthesis of delaminated graphene sheets by a proper choice of magnetic field modulating an electric-arc. An ultra-high purity glassy graphite anode was sublimated in an argon atmosphere. Carbon nanotubes (CNTs), as well as graphene sheets were found inside the deposit formed on the cathode. Both the high purity CNTs as well as graphene sheets, with minimal structural defects, were synthesized separately by varying the strength and orientation of the external magnetic field generated by arrays of permanent magnets. The as-synthesized carbonaceous samples were characterized with the help of transmission electron microscopy, selected area electron diffraction (SAED), Raman spectroscopy (RS) and thermogravimetry for optimizing the highest selective production of delaminated graphenes. This optimization was done by varying the strength and orientation of the external magnetic field.

The as-synthesized graphene sheets exhibited relatively high degree of graphitization and low structural defect density as confirmed by RS. They were found to exhibit higher oxidation temperature (767°C) than that of the carbon nanocrystalline (690°C) particles as inferred from the thermogravimatric analysis. Moreover, they were found to form 'scroll-like CNTs' at their edges on account of their surface energy minimization. This was confirmed by the SAED analysis. With this new technique, we could successfully synthesize delaminated graphenes at a rate of few g/h.

**1. Introduction**

The first historical isolation of single layer graphene from the crystalline graphite was demonstrated by Novoselov *et al* in 2004 [1]. Since then, graphene has snaffled huge scientific interests because of its unusual physical and electronic properties [2]. It has been demonstrated that inside graphenes, the electrons behave as mass less Dirac fermions [3] and a room temperature quantum Hall Effect can be realized within graphenes [4]. Moreover, these studies have demonstrated the importance of graphene as a 2D crystalline material for high electron mobility applications. However, the production methods of graphenes also suffer from the serious problem of lack of an eco-friendly way to produce the material at a bulk scale, commensurate with the application areas [5, 6]. There are currently three approaches to synthesize graphenes— mechanical or chemical exfoliation of pure graphite [1, 7], epitaxial growth by vapor phase deposition of hydrocarbons or CO on metal substrates [8-11] and thermal annealing of SiC substrates [12], each one of these involves the top down approach. In the current world scenario, chemical synthesis method has an edge due to its large scale synthesis potentials. However, the chemical routes are riddled with the issues of bio-compatibility. The present paper demonstrates the synthesis of high-quality graphenes directly from the gas phase condensation of carbon precursors, using a rotating electric-arc technique. The quality of the product is shown to contain low defects, wherein, the production rate is appreciably high (few g/h). The process is 100% eco-friendly.

Arc synthesis part of novel carbon structures, like, fullerenes [13] and carbon nanotubes (CNTs) [14], has been known to yield the best quality products [15]. Among various arc manipulation schemes available, magnetic field assisted electric-arc is an interesting area of research both for the plasma physicists and nanotechnologists [16-18]. Though, many reported works in the literature have addressed the effects of an external magnetic field on the formation of arc derived CNTs [19-21], none of them has addressed the following issues:
 (i) Effects on synthesis of carbon nanostructures when a carbon arc is superimposed with a steady non-uniform magnetic field with variable radial and axial components;
(ii) Feasibility of synthesizing a variety of carbon nanostructures (CNTs, graphene sheets etc.) through suitable combination of electric and magnetic fields;

(iii) Addressing suitable and effective ways to synthesize both CNTs and delaminated graphenes separately by parametric variation of a carbon arc.

The current study presents few significant insights to some of these issues through experiments on the magnetic field assisted electric-arc and the corresponding product analysis.

## 2. Experimental details

The experimental electric arc reactor, with arrangements to superimpose an external magnetic field (symmetrically about the arc axis) was designed and fabricated. After each operation, the as-synthesized carbon products were collected and thoroughly analyzed with a number of characterizing tools. The following sections deal with the experimental methodology adopted for this purpose.

*2.1. Description of the reactor*

The schematic of the electrode assembly with an external steady non-uniform magnetic field is shown in figure 1 (a). Here, the view of a vertical cross-section of this reactor along a plane passing through the axis of the reactor is shown. The main parts of this reactor consisted of two annular arrays of permanent ferrite magnets .The *upper* magnet array was aligned in a manner that all the south poles faced the axis of the array and their individual axes were collinear with that of the main electrodes. Both the magnet-arrays were arranged in eight fold symmetry about the reactor axis. The magnets were placed inside two annular brackets, made up of pure aluminium. Inside the upper bracket, south poles of the magnets faced the axis of the bracket, whereas, all the north poles faced the bracket-axis for the magnets placed inside the lower bracket.

The separation ($\Delta Z$) between the lower face of the upper array and the upper face of the lower array, as shown in figure 1(a), could be manually adjusted with a precision of 1mm. The values of $\Delta Z$ were adjusted symmetrically with respect to azimuthal plane. The anode was situated about 10mm away from the mid-plane as shown in the diagram. Due to the restrictions of the dimension inside the chamber, the value of $\Delta Z$ could be varied from a minimum of 20 mm to a maximum of 80 mm.

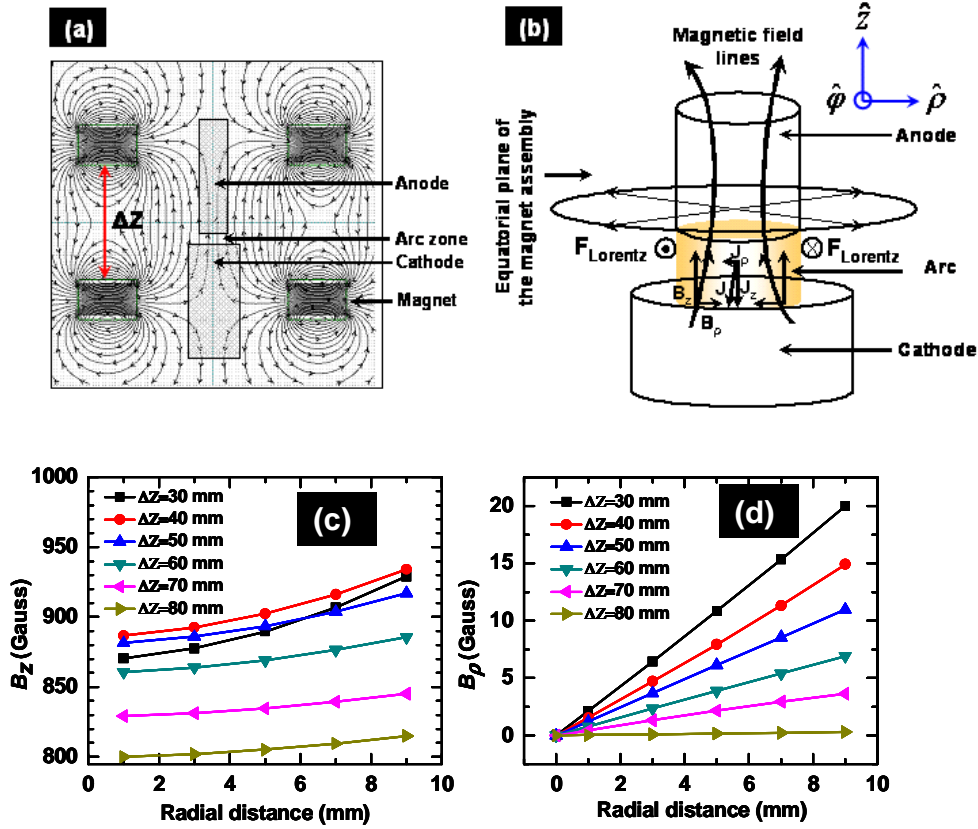

**Figure 1.** (a) Schematic of the vertical cross-section of the magnet assembly, (b) the distribution of the axial and radial components of external magnetic field and the current density vectors in the arc zone, (c) variation of the axial component of the magnetic field with ΔZ and (d) variation of radial magnetic field with ΔZ.

The dimensions of the magnet assembly are provided in table 1. This geometrical distribution of the permanent magnets could generate magnetic lines of force distributed symmetrically about the axis of the arc (figure 1(a)). This arrangement of the magnets provided a means to vary the strength as well as radial field distribution by varying ΔZ. Commercial ferrite magnets with the pole strength of 0.14T having Curie point of 170°C were used to produce the required magnetic field.

**Table 1.** Dimensions of the magnet assembly.

|     | Specification | Unit | Value |
| --- | --- | --- | --- |
| 1.  | Dimension of the pole faces of the magnets | mm$^2$ | 25×25 |
| 2.  | Width of each magnet | mm | 12 |
| 3.  | Effective magnetic pole strength | Gauss | 6.4k |
| 4.  | Radial spacing between two consecutive magnet blocks in an array | Degree | 45 |
| 5.  | Inner diameter of the magnet arrays | mm | 90 |
| 6.  | Outer diameter of the magnet arrays | mm | 162 |
| 7.  | Separation between the walls of the double-walled water cooling jacket | mm | 10 |
| 8.  | Accuracy of measuring ΔZ | mm | 1 |
| 9.  | Minimum value of ΔZ | mm | 20 |
| 10. | Maximum value of ΔZ | mm | 80 |

Water-cooling was provided (flow rate = 20 lpm) with the help of a double-walled stainless steel (non-magnetic) jacket for maintaining the temperature of the magnets below their Curie point. The entire assembly was mounted inside the chamber, the details of which were provided in our earlier communication [22].

In the present magnet assembly, both the axial and the radial magnetic field strengths could be varied by changing ΔZ. The strengths of both the axial and radial magnetic field were measured with the help of a Hall probe. The measured values are shown in figure 1(c, d). The magnetic field strengths, thus obtained, could regulate the strength of the Lorentz force, the description of which is provided in section 3.5.

The components of the magnetic field were measured in absence of the electrodes, arc and the water cooling jacket because of the space constraint and the experimental convenience. The estimated values are presented in figure 1 (c, d) to show the trend of their variation with ΔZ. However, it was not possible to correctly estimate their actual values during the experiments at the location of the arc.

*2.2. Synthesis methodology*

Initially, a base pressure of $10^{-3}$ Torr was obtained in the reactor chamber by an oil-rotary vacuum pump. The chamber was filled with argon and the operating pressure in the reactor chamber was maintained at 500 Torr by making use of a throttle valve. The experiments were run at a steady arc current of 170A (with a fluctuation of ±5A) and an arc voltage of 22V (with a fluctuation of ±2V) in an atmosphere of 99.99% pure argon. The arc was struck in between two 99.99% pure and solid cylindrical graphite (Graphite India Limited) rods with co-linear axes. One with 13 mm diameter served as an anode and was biased with the help of a current regulated power supply (6 kW capacity with an open circuit voltage of 60V). On the other hand, the other electrode with 30 mm diameter was grounded and served the purpose of cathode. The orientations of these electrodes have been shown in figure 1 (a).

The value of ΔZ was varied from 20 mm to 80 mm with a step of 10 mm. However, for maintaining a steady arc for all the values of ΔZ, the arc voltage was kept fixed at 22V. The operating conditions for this set of experiments are summarized in table 2.

After the completion of each synthesis run, the chamber was allowed to cool down to ambient temperature and the hard cylindrical deposits on the cathode surface were then dismantled for characterizations. The soft inner cores were not separated out and the cathode deposits (CDs) were characterized in their totality. The weight loss of the anode (with respect to its initial weight) and the corresponding weight of the cathode -deposit were measured after each synthesis run with the help of a sensitive digital balance (100 g capacity with a least count of $10^{-4}$ g). After collection, the hard deposits were mechanically homogenized with the help of pestle and mortar.

For the purpose of TEM analysis, the powdered samples were thoroughly dispersed in NN-dimethylformamide [H·CO·N(CH$_3$)$_2$] for about 0.5h. A small drop from each of these solutions, thus prepared, was then used for preparing the specimens on carbon coated copper grids. For the rest of the characterizations, the as-prepared powdered samples were employed with no further post synthesis treatment.

The TEM analysis was carried out by a 200 kV TECNAI G$^2$ 20 microscope, equipped with a LaB$_6$ filament and a CCD camera. Raman spectra were recorded in back-scattering geometry at room temperature with the help of a Jobin Yvon Labram HR800 spectrometer, in which a He–Ne laser ($\lambda$ = 632.81 nm) was used for the Raman excitations. The integration time for recording the Raman spectra was 100s and the spectra were averaged over three accumulations from the different parts of the same sample. The TG analysis was carried out in flowing dry air ambience with a flow rate of 8.33×10$^{-7}$ m$^3$ s$^{-1}$ by a 92-16.18 Setaram gravimeter at a temperature ramp of 5°Cmin$^{-1}$. For the TG analysis, the initial mass of each of the samples was taken to be about 8 mg.

**Table 2.** Specifications of the experimental parameters.

| | |
|---|---|
| Arc voltage | 22 V |
| Arc current | 170A |
| Diameter of the anode | 13mm |
| Diameter of the cathode | 30mm |
| Operating pressure | 500 Torr |
| Separation between the magnet arrays ($\Delta Z$) | 20mm to 80mm |
| Electrodes' composition | 99.99% pure graphite |
| Buffer gas | 99.99% pure Ar |

## 3. Results and Discussion

Initially, the arc was run under the conventional field free conditions. The magnet arrays were removed from the water-cooled jacket and the arc was found to be nearly steady. However, on imposing the external magnetic field, symmetrically (cylindrical symmetry) about the arc axis, the arc was found to gyrate on the cathode surface uniformly. The extent of gyration increased steadily with decrease in the value of $\Delta Z$. At $\Delta Z$=20mm, the arc became unstable and got extinguished frequently. The data corresponding to $\Delta Z$=20mm, were deemed unreliable and not analyzed in further detail.

The results revealed some of the new and interesting information on the mass of the CD, the evaporation from the anode, relative abundance of the various allotropic forms of carbon (including graphenes and CNTs) and Raman excitations due to the external magnetic field.

*3.1. Efficiency of formation of CD*

The CD formed in a carbon arc is the only source of nanocrystalline structures of carbon comprising of CNTs and CNPs [23].The rest of the deposits, which are found outside the arc zone, were mostly amorphous carbon with traces of the members of the fullerene family. The percentage conversion ($\eta_{CD}$), of the consumed anode material into CD, is therefore of practical importance for industrial applications. In case of the present reactor, the chosen operating parameters resulted into an average anode erosion rate of about 500mg/min.

Figure 2 shows the variation of $\eta_{CD}$ as a function of $\Delta Z$. It is seen that in absence of the magnetic field, about 90% of the consumed anode material is converted into CD. However, with decrease in the value of $\Delta Z$, this value decreases steadily reaching a minimum of about 55% at $\Delta Z$=60mm. With further increase to $\Delta Z$, the value of $\eta_{CD}$ is found to increase again reaching a value of about 60% at $\Delta Z$=30mm.

Earlier work by us had demonstrated a value of $\eta_{CD}$ *(optimum electric field conditions)* close to the maximum obtained here [22, 24]. However, in absence of any external field, the value of $\eta_{CD}$ was much less at 35% with Ar-He and He-H$_2$ mixtures as buffer gases [22, 24]. When the buffer gas was changed to helium, the value of $\eta_{CD}$ was found to be about 50% in absence of the external magnetic field.

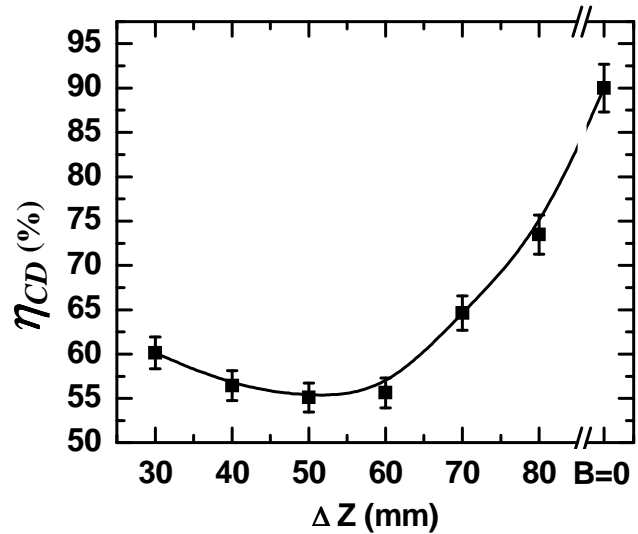

**Figure 2.** Variation of the % conversion efficiency of the consumed anode material into CD with ΔZ.

The data shows that the value of $\eta_{CD}$ is greatly influenced by the nature of the buffer gas and the heat radiating systems. It seems that on enveloping the arc with a heavier gas helps in achieving better value of $\eta_{CD}$. This finding is extremely significant as it provides another control variable for the efficient synthesis of nanocrystalline carbon structures. Though this result is in accordance with the study reported in the literature [25] the exact mechanism in the plasma, that governs the dependence of $\eta_{CD}$, is not understood well and can further be investigated in more detail in future scope of this work.

*3.2. Variation of the diameter of CD*

The magnetic field also distinctly affected the diameter of CDs ($\phi_{CD}$) for different values of ΔZ (Figure 3). It is seen from the figure that in absence of the magnetic field, $\phi_{CD}$ is very close to the diameter of the anode used.

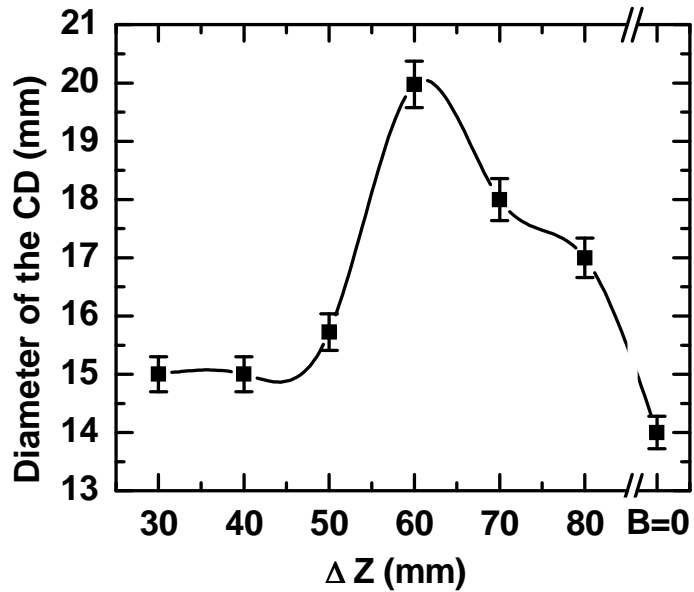

**Figure 3.** Variation of the diameter of CD as a function of ΔZ along with the one synthesized under no external magnetic field condition.

However on decreasing ΔZ, this value first steadily increases reaching a maximum of 20mm at ΔZ=60mm and then decreases steadily to a value of 15mm at ΔZ=30mm. This type of behavior was not observed earlier [22, 24]. The increase in the value of $\phi_{CD}$ is a clear signature of the spatial expansion of the plasma plume by the magnetic field and a decrease in the same corresponds to the confinement of the CD formation zone.

*3.3. Morphological analysis using TEM*

Having observed the trend in figure 2 and 3, a few of the as-synthesized CDs were analyzed with the help of TEM in order to find out their constituents' morphologies. Figure 7 highlights typical TEM micrographs of the sample prepared in absence of an external magnetic field. These TEM micrographs were recorded at different magnifications from different sites in the grid.

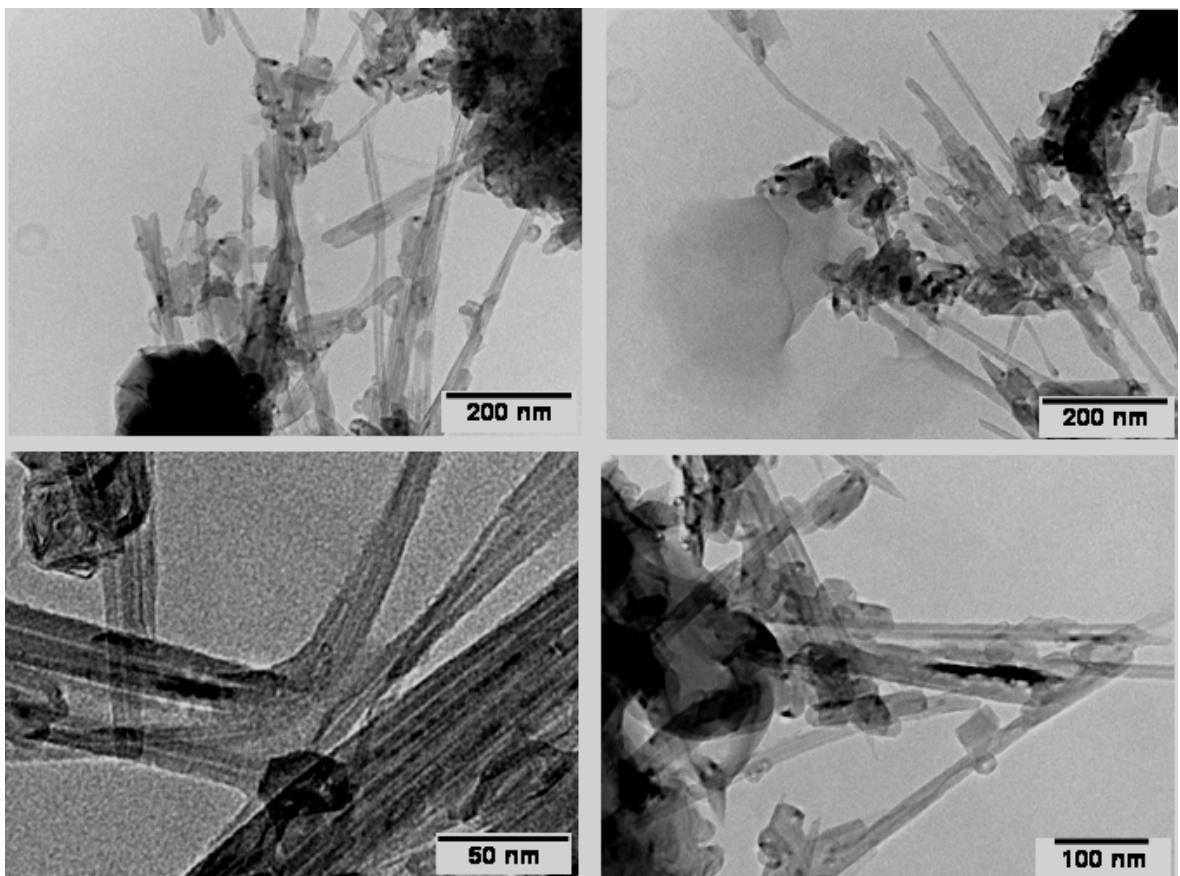

**Figure 4.** Typical transmission electron micrographs of the sample synthesized under zero magnetic field condition.

Figure 4 shows the presence of a large quantity of CNTs along with some CNPs with no traces of graphene like structures. The diameter distribution of the CNTs is observed to be wide, ranging from 20nm to more than 50nm, and most of the CNTs are seen to form bundles. The wall-thickness of each of these CNTs is indicative of their multi-walled nature.

Distinct structural transformations of the constituents of the CD generated in presence of the magnetic field were noticed in the corresponding TEM micrographs. Figure 5 shows some of the typical TEM micrographs of the sample generated at $\Delta Z=60$mm.

Distinct features of the micrographs are the graphene sheet like structures as seen in figure 5. Most of these sheets are seen to curl, on one side, to form the tubular structures. The tubular structures, thus formed, are seen to follow the typical 'scroll' model of MWNTs as has been highlighted by the curved arrow in figure 5. The red block-arrows in this figure, point out the areas where the graphitic sheets are rolled up. It seems that the CD formed at $\Delta Z=60$mm is an exuberant source of graphene like structures containing very few CNTs and CNPs.

Although, direct imaging of atomic structures is possible using high resolution electron microscopes, for determining an unknown crystal structure, or for making exact measurements of the structural parameters, it is necessary to rely on the diffraction experiments. The greater information content of such experiments lies in the fact that the diffraction process is optimally sensitive to the periodic nature of the solid's atomic structure.

In order to investigate the graphene like structures, synthesized at $\Delta Z=60$mm, in more detail, selected area electron diffraction (SAED) patterns were recorded from typically two regions, which are marked by yellow and blue circles in figure 5.

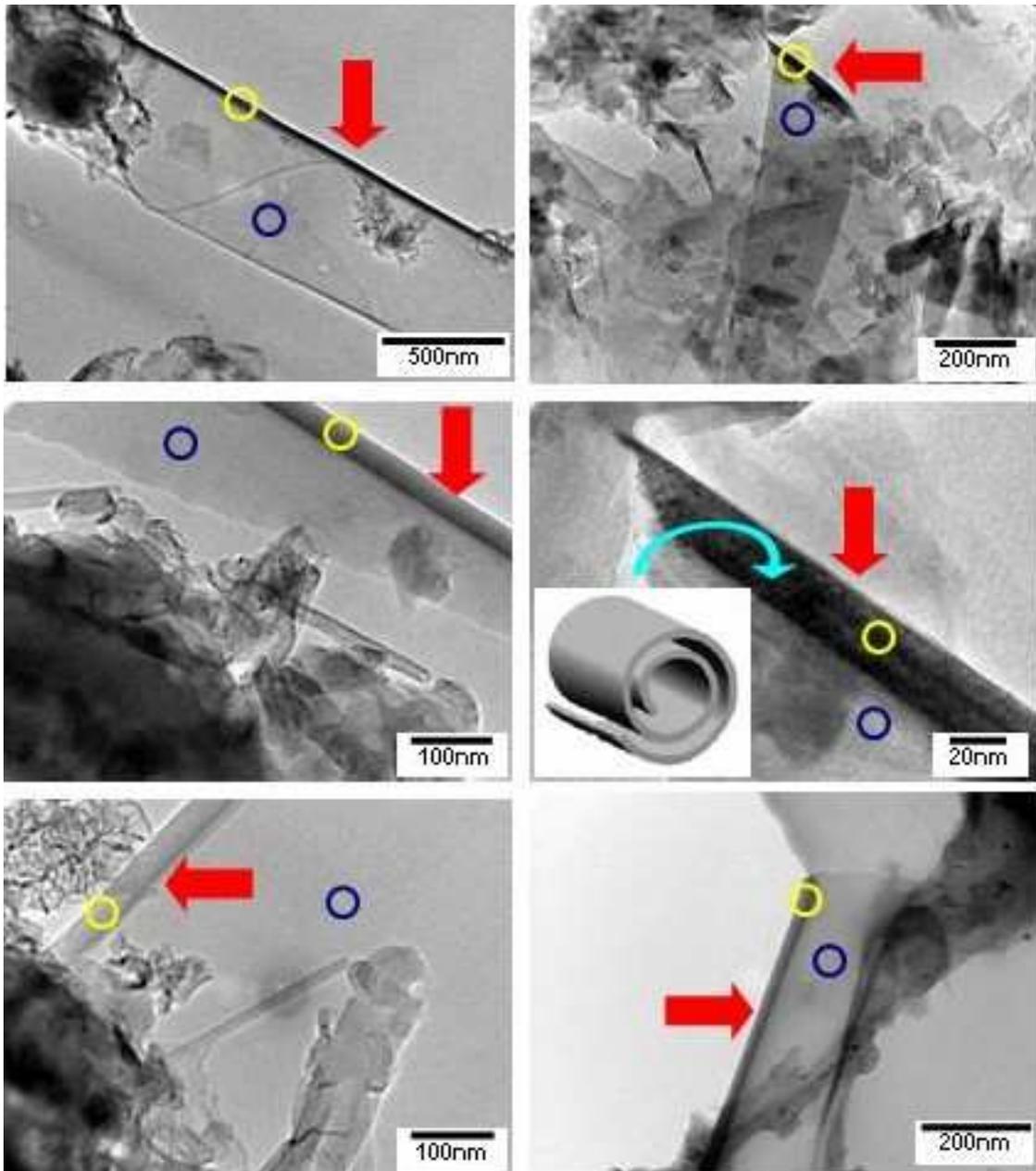

**Figure 5.** Typical transmission electron micrographs of the sample corresponding to ΔZ=60mm. The yellow circles correspond to the typical SAED pattern, which is shown in figure 6(a), while

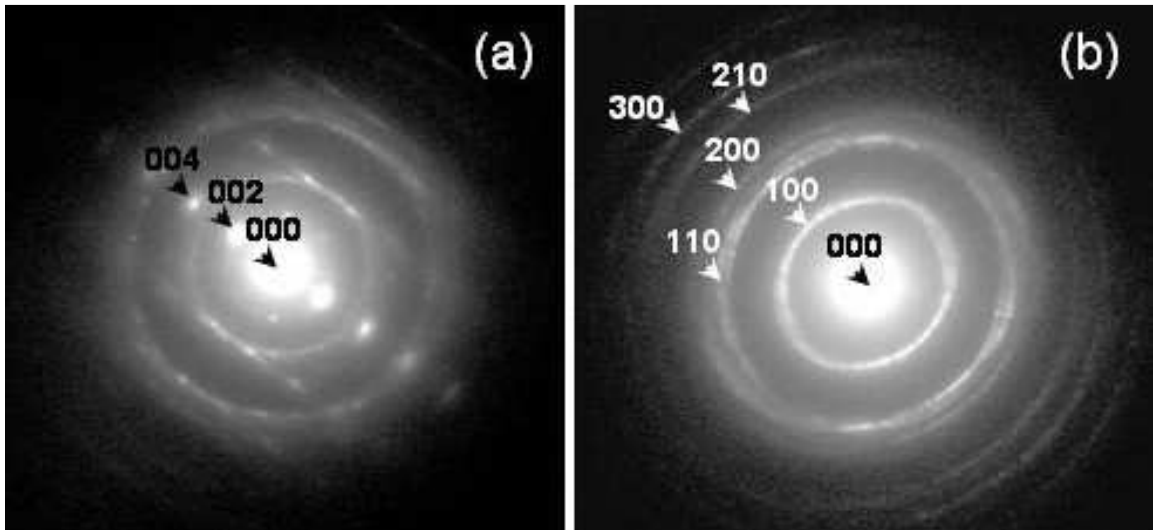

**Figure 6.** (a) Typical SAED pattern obtained from an area shown by yellow circles as indicated in Figure5.5 and (b) the same obtained from the green circles as indicated in the same figure.

the blue circles correspond to the typical SAED pattern as is shown in figure 6 (b).

The difference between these two SAED patterns is clearly visible. The characteristic features of the SAED pattern in figure 6 (a) are the presence of circular rings along with some symmetric bright spots. The rings correspond to (hk0) type reflections of graphite and the bright spots correspond to the lattice spacing of (00l) plans of graphite. The streaks of the intensity radiating outward and parallel to [00l] from (hk0) type rings approve the [14] independent scattering from delaminated curled graphitic sheets. These features of the electron diffraction pattern correspond to a typical electron diffraction taken perpendicular to a CNT, as reported in the literature [14].

On the other hand, the SAED pattern in figure 6 (b) is seen to comprise of oval shaped rings. In electron diffraction experiments we are usually accustomed of observing circular Debye rings. The occurrence of oval shaped nearly continuous rings is therefore a bit unusual. After measuring along the minor axes of the oval shaped rings, as indicated by arrows, we could successfully correlate these rings with (hk0) reflections (100), (110), (200), (210) and (300). The shape of the Debye rings from powder samples is exact circular, because; it appears as an intersection of the Ewald plane with the sphere of reflection in the reciprocal space. Thus, concentric oval shaped diffraction rings will appear when the Ewald-plane will intersect coaxial hollow cylinders of reflections in the reciprocal space. It is very obvious from diffraction results that only (hk0) type reciprocal lattice points of graphite have got extended in the reciprocal space. This extension is measured to be roughly $\sim 0.5 \text{Å}^{-1}$. This could be calculated taking into consideration of the difference between the lengths of the semi major and semi minor axes of the oval shaped diffraction pattern. Such a large 1D extension in reciprocal space directly implies that the layers of graphite have got delaminated to a thickness of few atomic orders, i.e it corresponds to graphene sheets. The lateral extents of these sheets are large. The presence of nearly continuous nature of oval rings dictates that (hk0) graphitic sheets are randomly oriented about an axis perpendicular to this.

Figure 7 shows typical TEM micrographs corresponding to ΔZ=30mm.

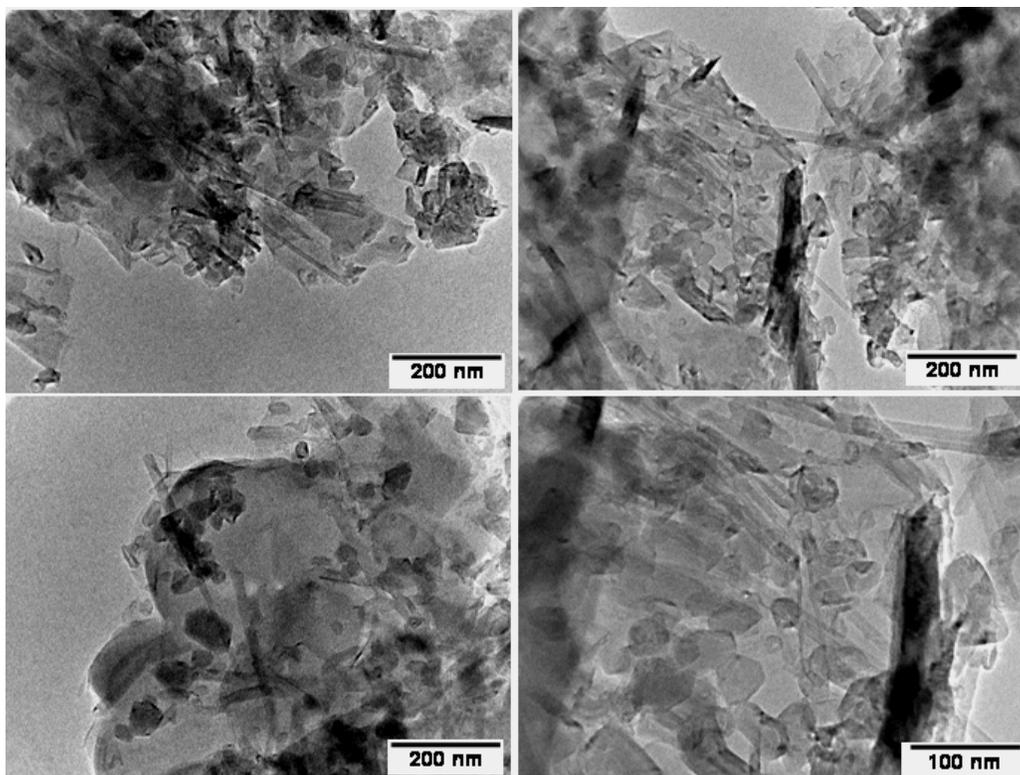

**Figure 7.** Typical transmission electron micrographs for the sample corresponding to ΔZ=30mm.

Distorted carbon structures, other than CNTs, are easily noticeable in these micrographs. The majority of the constituents of the corresponding CD seem to be the CNPs along with traces of few CNTs.

Analyzing figure 6 and figure 7 carefully; it is inferred that the external magnetic field, as has been used in the current study, does not help in the formation of CNTs in an arc process. It ceases the growth of CNTs and helps in the formation of graphene like structures. In other words, one can infer that the directional motions of the carbon precursors, which are thought to be responsible for the growth of CNTs [22] in the arc, are hampered in presence of an axial magnetic field. The situation rather promotes the formation of 2D structures in an arc process.

*3.4. Thermogravimetric analysis*

The oxidation behavior of the CDs is another important property and gives information about their crystalline purity and composition. While carrying out the TGA measurements, it was found that none of the samples underwent any weight-loss before 500°C. In order to get better accuracy and resolution, TGA measurements, therefore, were carried out from 500°C to 1000°C. Figure 8 summarizes the outcomes of this study. In the first column of figure 8, the percentage weight loss (TG) of the samples as a function of temperature for different values of ΔZ. The initial weight of the starting material is considered to be 100%. It is seen that as the sample temperature is increased beyond 500°C, most of the samples start to reduce their weight. This weight-loss is accounted for the formation of CO (g) and/ $CO_2$ (g) as a result of the oxidation of the samples by the $O_2$ molecules present in the air.

However, in order to understand the nature of oxidation for different samples in more detail, derivative of the TG (DTG) curves were obtained. The DTG curves are shown in the second column of figure 8 for different values of ΔZ. It is clearly seen from these curves that oxidation behavior are quite different from sample to sample. The oxidation of none of the samples, under investigation, is seen to be a single step process. This indicates that the samples undergoing oxidation are not composed of a single species. Rather, de-convolution of the DTG curves into a number of Lorentzian line-shapes, in order to obtain best fitting to the experimentally obtained DTG curves, clearly reveals the presence of a variety of components in different weight proportions. It is difficult to identify 'each and every' specie corresponding to each value of ΔZ due to the lack of exact knowledge of all these carbonaceous species.

However, oxidation temperatures of different carbonaceous species along with their contributions in weight percentage in each sample could be calculated with the help of the areas under the corresponding Lorentzian line-shapes. The results of these calculations are shown in the third column of figure 8. The distribution of oxidation temperatures and corresponding percentage weight contributions of different carbonaceous species as a function of ΔZ shows irregular behavior. This kind of fluctuation may best be attributed to the arc irregularities [26].

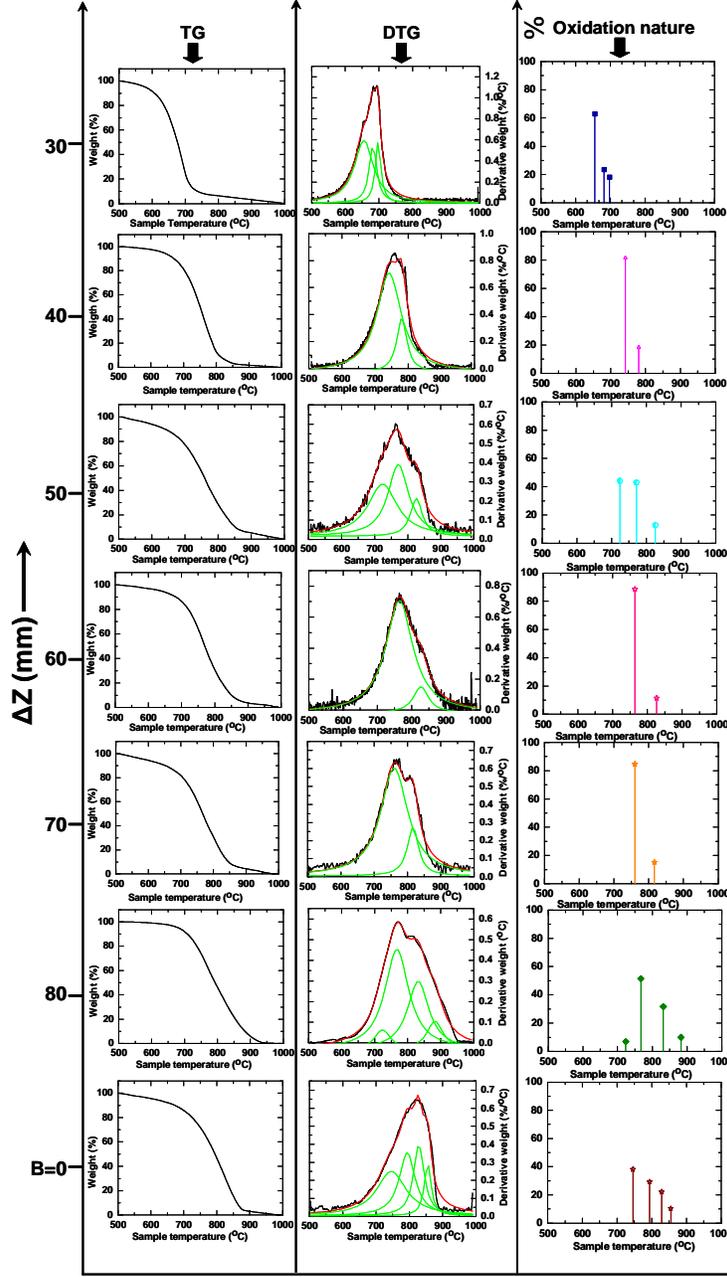

**Figure 8.** Oxidation behaviors of the composition of CD synthesized at different values of ΔZ along with the one synthesized under no external magnetic field condition. In the DTG curves, the green lines are the Lorentzian line-shapes and the red lines are the overall fitting.

The major impacts of the external magnetic field could be recognized, while looking at the most probable oxidation temperature ($T_P$) of different samples as a function of $\Delta Z$. $T_P$s were estimated from the DTG curves in figure 8 and correspond to the maximum amplitude of the associated curves. The variation of $T_P$ as a function of $\Delta Z$ is presented in figure 9. It is seen from this figure that the sample prepared in absence of the external magnetic field exhibits the highest value of $T_P$ (829°C). Though $T_P$ did not show much variation in the range of 40mm≤$\Delta Z$≤80mm, a closer look at the inset of figure 9 shows that, in this range of $\Delta Z$, there is a local minimum of $T_P$ (767°C) at $\Delta Z$=60mm. Moreover, the powdered CD generated at $\Delta Z$=30mm exhibited the lowest value of $T_P$ (690°C) among all the CDs synthesized in presence or absence of the external magnetic field.

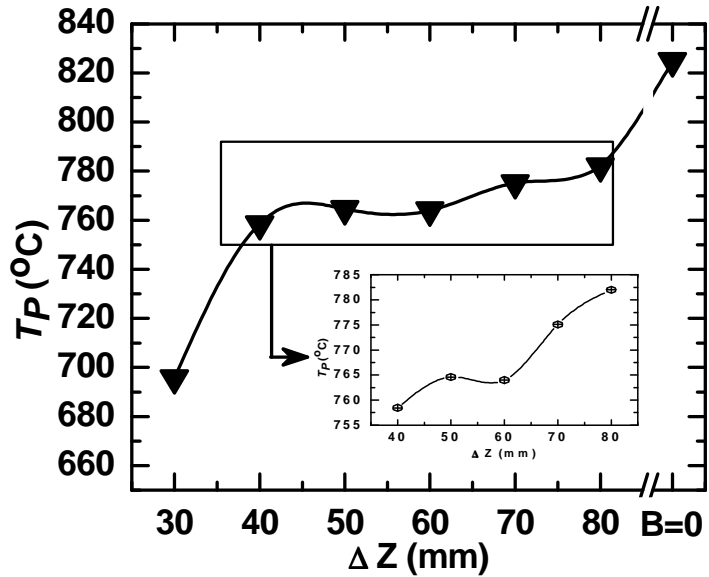

**Figure 9.** Variation of the most probable oxidation temperature of CDs synthesized at different values of $\Delta Z$ along with the one synthesized under no external magnetic field condition.

*3.4. Raman spectroscopic analysis*

RS analysis is a very powerful tool to infer the structural morphologies of different carbonaceous samples. It distinguishes the graphene structures from the other carbon structures and can provide significant information about the relative defect concentration and finite size effects in an unknown carbonaceous sample [27].

RS measurements were carried out in a slightly non-conventional way. Not only the overall quality of each sample was analyzed by this tool, but the CDs were also analyzed separately in more detail.

Firstly, for recording the Raman spectra, circular slices of about 5mm thickness were skillfully cut from each CD synthesized under different values of $\Delta Z$. Of these slices, three different portions were analyzed by RS. These positions were: (i) the center of CD, (ii) at R/2 position of CD, where R is the radius of the soft black core of the CD and (iii) at the edge of the soft black core of the CD as shown in figure 10. At least three different measurements were carried out for these three positions to identify the average features. A laser beam with a spot diameter of 1μm was used for this purpose.

In all the cases, RS were recorded in the range of 100cm$^{-1}$-1800cm$^{-1}$. However, no peak was identified in the lower range (100cm$^{-1}$-1100cm$^{-1}$) of the Raman spectra. These spectra exhibited the presence of only two distinct peaks, which could be identified as the D and G bands [27]. A typical such spectrum is shown in figure 11. Common features in all these Raman spectra are the three peaks. One, at around 1340 cm$^{-1}$, is correlated to the so-called disorder induced D band, and the other at around 1580 cm$^{-1}$ is referred to as the G band associated with a shoulder at 1620 cm$^{-1}$ [27]. To facilitate quantitative estimates, the relative intensities of G and D bands have been determined from the areas under the corresponding peaks. For enabling a proper comparison, all the G band amplitudes were normalized to unity. The G band intensities were determined after de-convoluting the G bands into two Lorentzians as shown in figure 13. $I_G$ and $I_D$ are the areas under the peaks at 1580 cm$^{-1}$ and 1330 cm$^{-1}$, respectively.

$I_G/I_D$ ratio was then calculated and plotted for the three different positions of the CDs generated under the different values of ΔZ, as shown in figure 12.

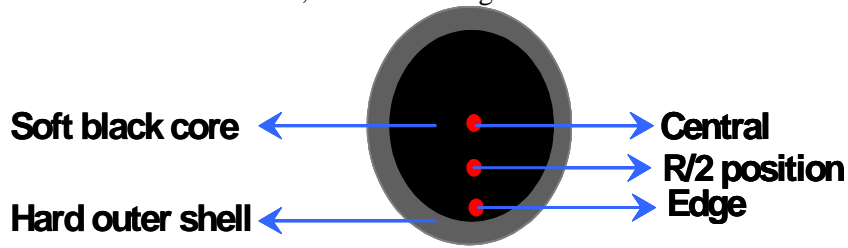

**Figure 10.** Schematic representation of the locations of the laser spot for in depth analysis of CDs by RS.

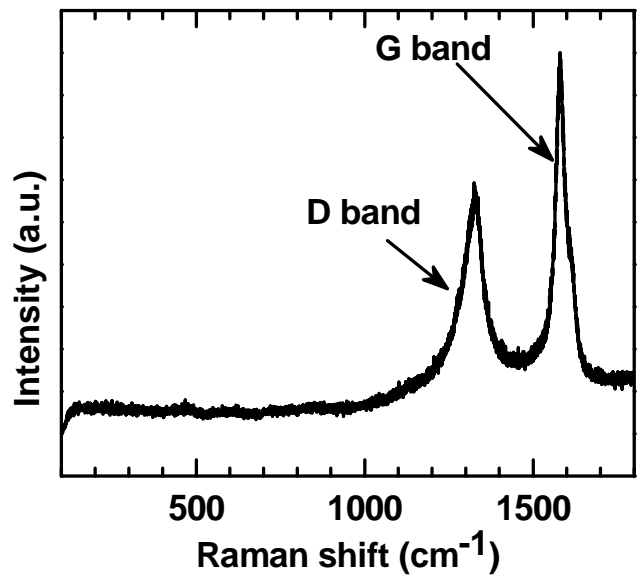

**Figure 11.** Typical first order Raman spectrum of the powdered CD.

It is seen from this figure that in absence of the external magnetic field, $I_G/I_D$ ratio exhibits a maximum at the center of corresponding CD and its value steadily decreases on moving towards the edge of the CD. On the other hand, the maxima of this ratio is seen to shift towards the edge of the CD in presence of the magnetic field.

The G band arises due to in-plane vibrations of the hexagonal graphene-like structures [27]. The broad hump centered at around 1340 cm$^{-1}$ (D band) has been extensively studied and has been explained by the relaxation or breakdown of the wave vector selection rules due to the finite size of the crystals in the bulk material [28] and a peak in the Raman spectra occurs near the maxima in the phonon density of states. In the case of our samples, the D band corresponds either to the CNPs due to their finite size or to the structural defects in the nanostructures, the relaxation of wave vector being equally valid for both.

The $I_G/I_D$ ratio is a well-accepted index for analyzing the fractional content of the ordered graphene structured species within a carbonaceous sample and an increase in this ratio indicates an increase in the graphene structured species and a decrease in the defect density, relative number of CNPs and percentage of a-C content [24].

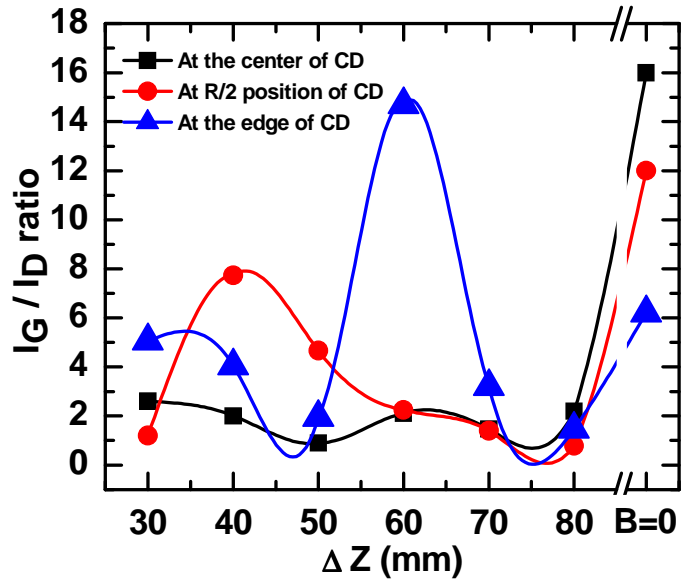

**Figure 12.** Radial distribution of the $I_G/I_D$ ratio for the CDs synthesized at different values of ΔZ along with the one synthesized under no external magnetic field condition.

The variation of the $I_G/I_D$ ratio in figure 12 is a clear indication of the fact that the central part of a conventional carbon arc is rich in graphene structured species. This is consistent with the fact that, CNTs are formed at the central part of the CD, thereby enhancing the overall graphene structured contents on account of the maximum local arc temperature at this position [23]. On moving towards the edge of the CD there is a gradual decrease in the local arc temperature, which promotes the formation of less graphitized species.

However, figure 12 reveals that on imposing the external magnetic field, the extent of graphitization shifts towards the periphery of the soft black inner core of the CD for 80mm≤ΔZ≤60mm. On decreasing the value of ΔZ below 60mm (increase in the radial magnetic field), the extent of graphitization again increases and shifts towards the R/2 position of CD. It is also equally noteworthy that the $I_G/I_D$ ratio remains almost equal at the center and the R/2 positions of the CDs corresponding to 80mm≤ΔZ≤60mm. This observation is a clear signature of the homogenization of radial distribution of local arc temperature possibly due to rotation of the arc column due to the magnetic field. It is also noteworthy that the nanostructures generated near the axial region of the arc in presence of the external magnetic field are always less graphitized than those generated away from the arc-axis. This trend is exactly opposite to that observed under no magnetic field condition.

Following these measurements, in order to check the overall quality of each sample, all the CDs were crushed thoroughly and Raman measurements were carried out again for these powdered samples.

The first column of figure 13 shows the first order Raman spectra recorded in the range of 1100cm$^{-1}$-1700cm$^{-1}$ for the mechanically homogenized CDs generated at different values of ΔZ. The presence of both the D and G bands is clearly seen in this figure.

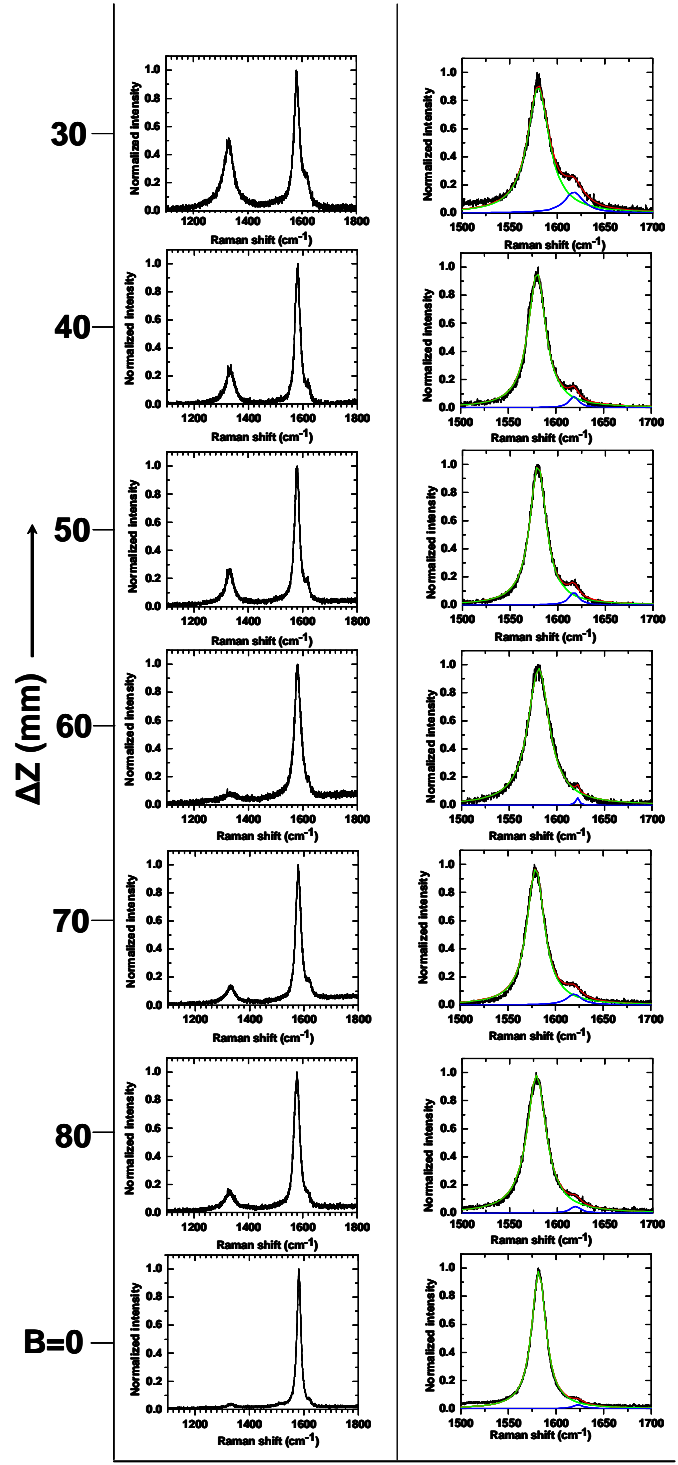

**Figure 13.** First column shows the first order Raman spectra of the mechanically homogenized CDs synthesized under different values of ΔZ along with the one corresponding to no external magnetic field. The second column shows the de-convolutions of the G bands into two Lorentzian line shapes.

In order to estimate an average relative abundance of the graphene structured species within the samples, $I_G/I_D$ ratio for each of the as-synthesized samples was calculated from the areas under the G and D bands. For these calculations, only the areas of the peak at 1582 cm$^{-1}$ was taken into account as the G band intensities. The intensities of the shoulder peak at 1620 cm$^{-1}$ were excluded from these calculations. Figure 14 shows the variation $I_G/I_D$ ratio as a function of ΔZ along with the one obtained for no external magnetic field condition.

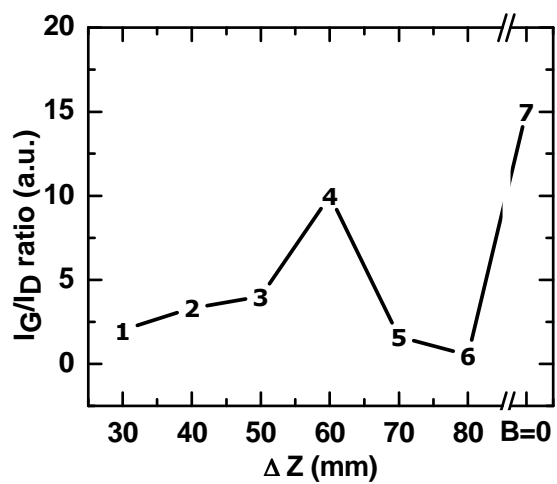

**Figure 14.** Variation of $I_G/I_D$ ratio for the mechanically homogenized CDs with ΔZ along with the one corresponding to no external magnetic field.

It can easily be noticed from this figure that point 7 exhibits the highest value of $I_G/I_D$ ratio with $I_G/I_D$=14. However, point 4 also exhibits a distinct maxima with $I_G/I_D$=10 among all the samples synthesized in presence of the external magnetic field. The trend of variation of $I_G/I_D$ ratio with ΔZ is very similar to that as was seen in the variation of the diameter of the CDs (figure 3).

From the values of these $I_G/I_D$ ratios, it is logical to infer that the sample synthesized under no external magnetic field condition, has the highest degree of graphitization and contains the least amount of structural defects. On imposing the external magnetic field, the extent of graphitization decreases. However, the sample synthesized at ΔZ =60mm is definitely richest in terms of the graphene structured contents among all the samples synthesized in presence of the external magnetic field.

A clear understanding of this behavior has not been achieved. However, an attempt at understanding the behavior of arc in presence of the used magnetic field assembly is attempted in the following section.

*3.5. Behavior of electric-arc in presence of the magnet assembly*

An electric-arc column at a pressure of 500 Torr, behaves like a pure electrical conductor. As the charged plasma precursors are in a high-collision regime, the effects of the external magnetic field will not be on individual precursors; rather the magnetic field will modulate the entire arc column. A typical nature of the magnetic field produced near the arc zone in the present case has been schematically shown in figure 1. It has also been found that magnetic field generated by the

magnet arrays is not perfectly axial. The magnetic field vectors, thus generated, can best be described as

$$\vec{B}(\rho, z, \Delta z) = B_z(\rho, z, \Delta z)\hat{z} + (-B_\rho(\rho, z, \Delta z))\hat{\rho} \quad \ldots\ldots\ldots\ldots (1)$$

taking into consideration the cylindrical symmetry of the arc. Here, $B_z$ and $B_\rho$ are the axial and radial components of the magnetic field respectively. The drift velocity of the plasma precursors can be expressed as

$$\vec{V}_d = V_z(\rho, \Delta z)\hat{z} + V_\rho(\rho, \Delta z)\hat{\rho} \quad \ldots\ldots\ldots\ldots\ldots\ldots\ldots\ldots\ldots\ldots\ldots (2)$$

for the present case. $V_\rho$ will always be positive, because, the dimension of the cathode is larger than the anode in the present case. As a result, the Lorentz force acting on a charged particle will take the form:

$$\vec{F}_{Lorentz} = q \times (\vec{V}_d \times \vec{B}(\rho, z, \Delta z)) = q[(V_z(\rho, \Delta z)B_\rho(\rho, z, \Delta z) - V_\rho(\rho, \Delta z)B_z(\rho, z, \Delta z)]\hat{\varphi}$$
$$= [J_z(\rho, \Delta z)B_\rho(\rho, z, \Delta Z) - J_\rho(\rho, \Delta z)B_z(\rho, z, \Delta Z)]\hat{\varphi} \quad \ldots\ldots\ldots\ldots\ldots\ldots\ldots (3)$$

making use of equations 1 and 2. Here, $q$ is the charge of the plasma precursor and $J$ is the current density. For a current carrying plasma column equation (3) would describe the net $\vec{J} \times \vec{B}$ force. So, Depending on the magnitude of the Lorentz force, the entire current conducting channel of the arc will gyrate in the $\pm\hat{\varphi}$ direction with different rotational frequencies following equation 3.

The dynamics of the arc column is decided by the magnitudes of $J_\rho$, $J_z$, $B_\rho$ and $B_z$ on the cathode and anode surfaces. As the current distribution at cathode and anode surfaces will critically depend on the surface-geometry of both the anode and the CD, they would decide the relative magnitudes of $J_\rho$ and $Jz$. At high values of $\Delta Z$; the magnetic field at the cathode is mostly axial (figure 1(c, d)). In the full range of $\Delta Z$, in the experiment, the ratio of axial to radial field varied between 0–100 at r=10 mm, where r is the radial distance close to the cathode surface (figure 1). Therefore, the net force would critically depend on $J_\rho$ and $J_z$ at the cathode.

As seen from the data on estimated CD diameter (figure 3), $\Delta Z$ =60 mm gives the maximum size of 20 mm. The diameter reduces as $\Delta Z$ is reduced below 60 mm and remains more or less constant below 40 mm. Similarly, the RS data indicated increased presence of graphene structured species at the periphery at $\Delta Z$ =60 mm whereas $\Delta Z$ =40 shows maximum graphene structures at r =R/2 and $\Delta Z$=30 shows high quantity of the graphene structured species again at the periphery. The thermogravimetry data indicated monotonic decrease in the most probable oxidation temperature, with a local minima at $\Delta Z$ =60 mm, as the magnetic field increased. On the other hand, the TEM features (figures 4, 5 and 7) of the CD compositions show transitions from CNPs to CNTs through graphene sheets formation on gradually withdrawing the external magnetic field.

As a result of the Lorentz force, the arc column will rotate about the axis of the electrodes and the cathodic spot will de-centralize and anchor towards the peripheral positions of the cathode. In consequence to this, the local arc temperature of the arc will be more towards the periphery of the CD than that at its central zone. The local graphitization will therefore be more towards the periphery of the CD. More the Lorentz force, more will extent of de-centralization of graphitizing zone. In view of this fact, it seems that the arc rotation is maximum at $\Delta Z$=60mm and decreases gradually on deviating from this condition. Figure 3 provides strong support to this conjecture.

Though it is extremely difficult to reconcile all the observed data through arc dynamics without detailed measurements of current density distribution, magnetic fields and associated simulation, $\Delta Z$=60 mm indicates a situation conducive to the enhanced formation of delaminated graphene sheets. A few remarks to explain the observed trend can be made as follows.

At first glance, the rotation of the arc column and the relative magnitude of the two competing terms $J_z B_\rho$ *versus* $J_\rho B_z$ appear to be responsible for the said behavior. This is further supported by the fact that graphitization is possible only at high temperature of the cathode surface. Possibly, at $\Delta Z = 60$ mm, the conditions are most favorable for the rotating arc column to cover a major area of the cathode. As $\Delta Z$ is decreased further, beyond 60mm, the force balance (equation 3) changes and the graphitization as well as the CD shrink towards the axis.

However, this simplistic scenario needs to be supported by detailed quantitative measurements and simulation of the current density vectors, plasma column diameter and speed of arc rotation. The problem is out of the scope of the present work and can be tackled as the future scope of this work.

## 4. Summary and Conclusions

The study of magnetic field assisted synthesis of carbon nanostructures in a carbon arc has resulted in extremely interesting data not reported in literature so far. It has been shown that arc electric field and steady non-uniform magnetic field can influence the precursor trajectories and energies to effect stacking of carbon atoms promoting the growth of CNTs or graphene sheet like structures. The conclusions drawn from the present study can be surmised as follows.

(i) The CD generated in the absence of an external magnetic field exhibited a high $I_G/I_D$ ratio of 14 in the Raman spectra. The corresponding most probable oxidation temperature was found to be quite high (829°C) (figure 13). These features support the fact that the corresponding CD is rich in MWNTs (figure 4).

(ii) An arc-plasma, under the action of a specially configured magnetic field, helps in the formation of graphene sheet like structures. The CD generated at $\Delta Z=60$mm exhibited a relatively high oxidation temperature (767°C) and it was found to be an exuberant source of graphene sheet like structures (figure 5 and figure 13). On the other hand, from figure 7 and figure 13 it is seen that the CD synthesized at $\Delta Z=30$mm was rich in CNPs and it showed a relatively low oxidation temperature (690°C). The graphene sheet like structures can be viewed as metastable or intermediate forms of CNT-growth. These sheets are highly graphitized and are expected to burn at lower oxidation temperature than CNTs and CNPs, which are known as the stable structures of carbon. However, on account of the observed higher oxidation temperature of the CD, rich in these sheets, with respect to the one rich in CNPs, it is possible that, while recording the TG spectra of the corresponding CDs, these sheets first roll up to minimize their surface energy and then get oxidized.

(iii) The behavior observed, below the optimum magnetic ring separation ($\Delta Z=60$mm), is not understood well. The corresponding CDs are found to be composed of mostly CNPs along with some traces of CNTs (figure 7). This might be due to plasma instability on account of imposition of the magnetic field with high strength.

However, the most significant contribution of present paper is unfolding a novel physical way, which is capable of producing large quantity of delaminated graphenes at a much faster rate directly from the gas phase condensation of carbon precursors in an electric arc method. The attempt is the first of its kind and is envisioned to attract more physicists to contribute significantly in the field of graphene monolayer synthesis, which is rather the choice of chemists at present [29].

**Acknowledgements**
The work was carried out under the scope of BARC-PU Joint Collaborative Research Program. The financial support from BARC, Mumbai and Board of Research in Nuclear Sciences under the Department of Atomic Energy, Government of India is gratefully acknowledged. SK specially acknowledges the Head, Chemistry Division, BARC, Mumbai, India for giving permission to carry out the thermogravimetric measurements.